\documentclass[preprint,prb,aps,showpacs,footinbib,amsmath,amssymb,superscriptaddress,longbibliography]{revtex4-1}
\usepackage{graphicx}
\usepackage{dcolumn}
\usepackage{color}
\usepackage{bm}
\usepackage{amsmath,amssymb,amsfonts,bbold}
\usepackage[normalem]{ulem}
\usepackage{upgreek}

\newcommand{\beq}{\begin{equation}}
\newcommand{\eeq}{\end{equation}}
\newcommand{\beqa}{\begin{eqnarray}}
\newcommand{\eeqa}{\end{eqnarray}}

\begin{document}

\title{Indirect Chiral Magnetic Exchange\\ through Dzyaloshinskii-Moriya--Enhanced 
RKKY Interactions\\ in Manganese Oxide Chains on Ir(100)}

\author{Martin Schmitt}
\email{maschmitt@physik.uni-wuerzburg.de} 
	\affiliation{Physikalisches Institut, Experimentelle Physik II, 
	Universit\"{a}t W\"{u}rzburg, Am Hubland, 97074 W\"{u}rzburg, Germany}
\author{Paolo Moras} 
	\affiliation{Istituto di Struttura della Materia-CNR (ISM-CNR), Trieste, Italy}
\author{Gustav Bihlmayer}
	\affiliation{Peter Gr\"{u}nberg Institut and  Institute for Advanced Simulation, 
                        Forschungszentrum J\"{u}lich \& JARA, 52425 J\"{u}lich, Germany}
\author{Ryan Cotsakis}
	\affiliation{Physikalisches Institut, Experimentelle Physik II, 
	Universit\"{a}t W\"{u}rzburg, Am Hubland, 97074 W\"{u}rzburg, Germany} 
	\affiliation{University of British Columbia, 2329 West Mall, Vancouver, BC 	Canada}
\author{Matthias~Vogt} 
	\affiliation{Physikalisches Institut, Experimentelle Physik II, 
	Universit\"{a}t W\"{u}rzburg, Am Hubland, 97074 W\"{u}rzburg, Germany}
\author{Jeannette~Kemmer} 
	\altaffiliation{Current address: California Institute of Technology, 1200 E.\ California Blvd., 
		Pasadena, CA, 91125 USA}
	\affiliation{Physikalisches Institut, Experimentelle Physik II, 
	Universit\"{a}t W\"{u}rzburg, Am Hubland, 97074 W\"{u}rzburg, Germany}
\author{Abderrezak Belabbes}
	\affiliation{Physical Science and Engineering Division,
	King Abdullah University of Science \& Technology (KAUST), Thuwal 23955-6900, Saudi Arabia}
\author{Polina~M.~Sheverdyaeva} 
	\affiliation{Istituto di Struttura della Materia-CNR (ISM-CNR), Trieste, Italy}
\author{Asish~K.~Kundu} 
	\affiliation{International Center for Theoretical Physics (ICTP), Trieste, Italy}
\author{Carlo Carbone} 
	\affiliation{Istituto di Struttura della Materia-CNR (ISM-CNR), Trieste, Italy}	
\author{Stefan Bl\"{u}gel}
	\affiliation{Peter Gr\"{u}nberg Institut and  Institute for Advanced Simulation, 
                        Forschungszentrum J\"{u}lich \& JARA, 52425 J\"{u}lich, Germany}
\author{Matthias Bode} 
	\affiliation{Physikalisches Institut, Experimentelle Physik II, 
	Universit\"{a}t W\"{u}rzburg, Am Hubland, 97074 W\"{u}rzburg, Germany}	
	\affiliation{Wilhelm Conrad R{\"o}ntgen-Center for Complex Material Systems (RCCM), 
	Universit\"{a}t W\"{u}rzburg, Am Hubland, 97074 W\"{u}rzburg, Germany}    
	   


\date{\today}

\pacs{}


\maketitle

{\bf Localized electron spins can couple magnetically via the 
Ruderman-Kittel-Kasuya-Yosida interaction even if their wave functions lack direct overlap. 
Theory predicts that spin-orbit scattering leads to a Dzyaloshinskii-Moriya type enhancement 
of this indirect exchange interaction, giving rise to chiral exchange terms.
Here we present a combined spin-polarized scanning tunneling microscopy, 
angle-resolved photoemission, and density functional theory study of MnO$_2$ chains on Ir(100). 
Whereas we find antiferromagnetic Mn--Mn coupling along the chain, 
the inter-chain coupling across the non-magnetic Ir substrate turns out to be chiral 
with a $120^{\circ}$ rotation between adjacent MnO$_2$ chains.  
Calculations reveal that the Dzyaloshinskii-Moriya interaction results in spin spirals 
with a periodicity in agreement with experiment. 
Our findings confirm the existence of indirect chiral magnetic exchange, 
potentially giving rise to exotic phenomena, such as chiral spin-liquid states in spin ice systems 
or the emergence of new quasiparticles.  
}  


The concept of the Ruderman-Kittel-Kasuya-Yosida (RKKY) interaction \cite{RK1954,Kas1956} has successfully been applied 
to explain the magnetic properties of numerous indirectly coupled material systems 
which cannot be properly described by direct Heisenberg exchange.
Prominent examples are the rare-earth metals 
with their partially filled but highly localized $4f$ shell \cite{RZK1966,BC1992} 
or magnetic multilayers separated by non-magnetic metallic spacers. 
Since spin-orbit--related effects play no significant role in conventional RKKY,
practical realizations are largely limited to collinear coupling terms, 
where---depending on spacer thickness and Fermi wave length---%
the relative magnetic orientation is either parallel or antiparallel \cite{Sti1999}.  
Nevertheless, the giant magneto-resistance effect of layered magnetic materials 
is widely used in spin valve applications for field sensors or magnetic read heads \cite{Har2000}. 

Theory predicts that spin-orbit scattering leads to a Dzyaloshinskii-Moriya \cite{Dzy1957,Mor1960} 
type enhancement of the RKKY interaction \cite{Smi1976,FL1980}, or DME-RKKY in short, giving rise to chiral exchange terms. 
First evidence of indirect chiral magnetic exchange in layered structures was obtained from 
magnetic field-dependent neutron diffraction studies Dy/Y superlattices \cite{GCY2008,GLC2010}. 
Further experimental evidence of DME-RKKY is essentially limited to non-collinear spin structures 
observed in surface-deposited clusters \cite{2015Dupe,Khajetoorians2016,Bouaziz2017,Hermenau2017}.    
Here we report on the direct observation of chiral magnetic order between MnO$_2$ chains 
which is mediated by RKKY interaction via conduction electrons of the Ir substrate.  
The strong spin--orbit coupling in Ir leads to an appreciable DMI, 
resulting in a chiral spin spiral with a $120^{\circ}$ rotation between adjacent MnO$_2$ chains. 

\section*{Results}
{\bf Structural and electronic properties.}
The growth and structural properties of self-organized transition metal oxide (TMO) chains on Ir(001) 
have recently been studied by means of STM and low-energy electron diffraction (LEED) \cite{2016Ferstl}.
It has been shown that many TMOs form extended $(3\,\times\,1)$-ordered domains.
Depending on the particular transition metal element, various intra-chain spin structures (along the chain)
were predicted by DFT calculations \cite{2016Ferstl}, 
ranging from a non-magnetic NiO$_2$, over ferromagnetically ordered (FM) CoO$_2$, 
to antiferromagnetic (AFM) FeO$_2$ and MnO$_2$ chains.
In contrast, only a very weak inter-chain magnetic coupling between adjacent chains 
across the Ir(001) substrate was predicted \cite{2016Ferstl}, 
too weak to result in spontaneous, permanent, and long-range magnetic order.  



\begin{figure}[t]   
\centering
\vspace{-1cm}\includegraphics[width=0.59\columnwidth]{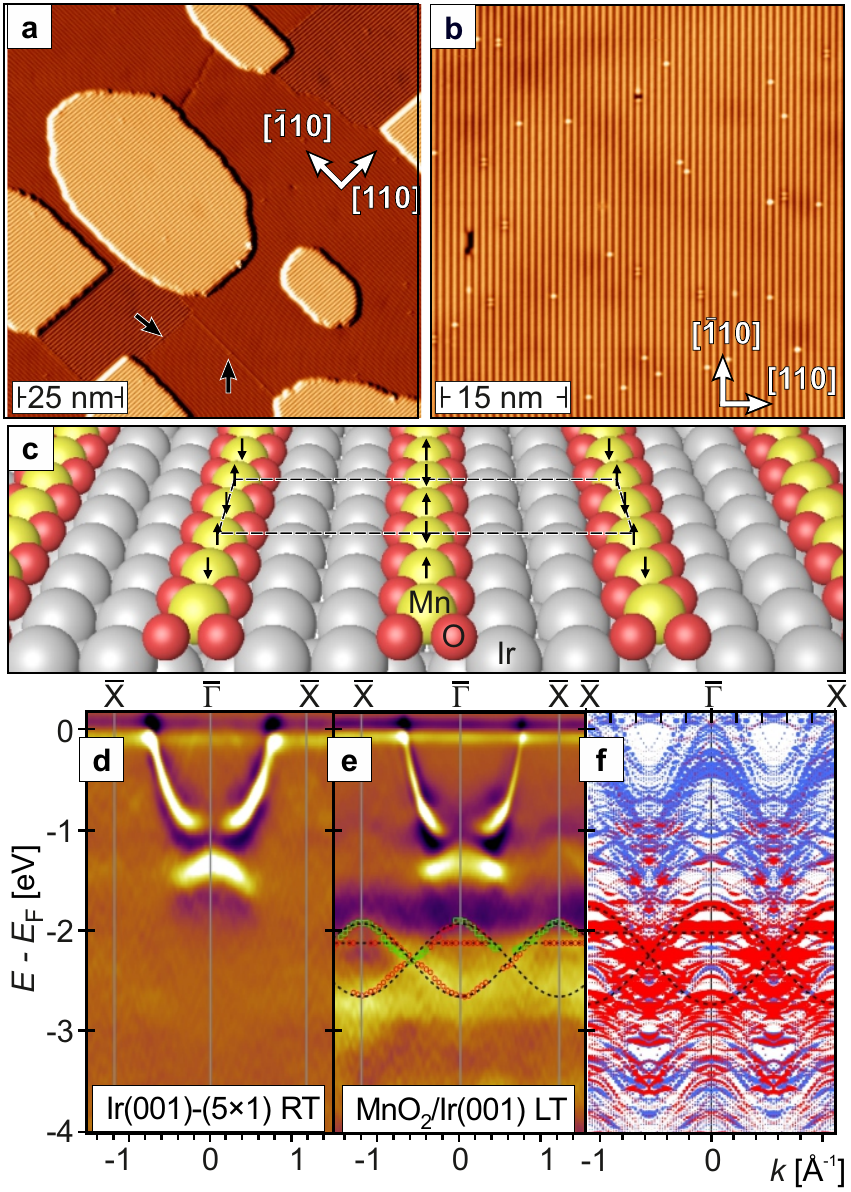}%
\vspace{-0.5cm}\caption{{\bf Structural and electronic properties of MnO$_2$ on Ir(001).} 
	{\bf a}, Large scale STM image showing islands of monolayer height. 
	The entire surface including the islands are covered 
	by chains along the $[110]$ or $[\overline{1}10]$ direction.   
	{\bf b}, Higher resolution STM image of the MnO$_2$ chains.
	Scan parameters: $U = 1$\,V, $I = 300$\,pA.
	{\bf c}, Schematic model of the atomic structure and the AFM 
	$(6\,\times\,2)$ spin arrangement predicted in Ref.\,\onlinecite{2016Ferstl}.
	{\bf d},{\bf e}, Second derivative ARPES data of the Ir(001)-$(5\,\times\,1)$ surface (at room temperature, RT) 
	and for MnO$_2$/Ir(001) (at 13 K, LT) along the $\overline{\Gamma}$--$\overline{X}$ direction ($h\nu$ = 150 eV). 
	Red and green symbols mark the dispersion of Mn-related states 
	as de\-ter\-mined from data taken at $h\nu$ = 150 eV and 130 eV, respectively.  
	Their periodicity is in agreement with the $2\times$ AFM order expected along the chains. 
	Dashed lines are guides to the eye. 
	{\bf f}, DFT band structure of MnO$_2$/Ir(001) exhibiting AFM order along the chains. 
	Red and blue dots represent Mn and Ir states, respectively . 
	The size of the symbols indicates the surface localization of the corresponding state. 
	The same dashed lines shown in {\bf e} (stretched by a factor 1.33 
	to consider the larger band width in DFT) agree well with surface-localized Mn states.}
\label{Fig:overview}
\end{figure}

A topographic STM image of a typical MnO$_2$/Ir(001) surface is shown in Fig.\,\ref{Fig:overview}a.
Wide flat terraces are decorated by roughly rectangularly shaped islands of atomic height. 
Terraces and islands both exhibit stripes running along the $[110]$ or the $[\overline{1}10]$ direction.  
These stripes originate from the self-organized growth of MnO$_2$ chains
which leads to a $(3\,\times\,1)$ structural unit cell \cite{2016Ferstl}.  
Some domain boundaries can be recognized which separate domains which differ either in stripe direction 
(left arrow in Fig.\,\ref{Fig:overview}a)
or by an incommensurate phase shift (right arrow).
The higher resolution image of Fig.\,\ref{Fig:overview}b 
was measured on a single $(3\,\times\,1)$ domain. 
The stripe periodicity of $(840 \pm 50)$\,pm,  
corresponds well to the expected value of $3 \times a_{\rm Ir} = 816$\,pm, 
with the Ir lattice constant $a_{\rm Ir} = 272$\,pm \cite{2016Ferstl}.
Only 36 defects are observed 
(24 bright spots; 9 dumbbells, 2 point-like hole; 1 line defect), 
equivalent to a chain defect density below 0.35\%. 

The structure of the MnO$_2$ chains on Ir(001) as proposed by Ferstl {\em et al.}\,\cite{2016Ferstl} 
is schematically represented in Fig.\,\ref{Fig:overview}c.  
Along the chains nearest-neighbor Mn atoms (yellow) are separated by two oxygen atoms (red).
Interestingly, the MnO$_2$ chains sit above empty substrate rows, held in place by the oxygens atoms.
DFT calculations predicted an AFM coupling along the MnO$_2$ chains, 
favored by 27\,meV per Mn pair with respect to a FM coupling \cite{2016Ferstl}.  
Due to the large separation between adjacent chains a much weaker AFM coupling 
with an energy gain of 0.4\,meV per Mn pair was found across the stripes, 
overall resulting in a rectangular $(6\,\times\,2)$ magnetic unit cell, sketched in Fig.\,\ref{Fig:overview}c.
ARPES measurements support the presence of an AFM intra-chain coupling. 
Fig.\,\ref{Fig:overview}(d,e) display second derivative ARPES spectra along the $\overline{\Gamma}$--$\overline{X}$ axis 
of the Ir(001) surface Brillouin zone (SBZ) (corresponding to the Ir [110] direction) 
for the clean $(5\,\times\,1)$-reconstructed substrate and the MnO$_2$/Ir(001) system, respectively. 
The second derivative is used to enhance the sensitivity to Mn-related states, which are broadened by the hybridization with the substrate.

The photon energy is chosen such that the Ir $5d$ signal is weak, 
except for some bulk bands dispersing symmetrically about $\overline{\Gamma}$ within 1.6\,eV below $E_{\rm F}$. 
Upon formation of the MnO$_2$ chains new states appear between $-1.9$ and $-2.9$\,eV (Fig.\,\ref{Fig:overview}e). 
The peak positions of Mn-related states are marked by red ($h\nu$ = 150 eV) and green symbols ($h\nu$ = 130 eV). 
Two sinusoidal dashed lines having maxima and minima at the $\overline{\Gamma}$ 
and $\overline{X}$ points are guides to the eye connecting the dispersive states. 
Flat states below the maxima are connected by dashed segments (see Supplementary Note 1 for bare data). 
These lines are compared with first-principles electronic structure calculations 
of AFM MnO$_2$ chains on Ir(001) oriented along the $y$-direction (Fig.\,\ref{Fig:overview}f). 
They match well with the energy position of surface-localized Mn bands, but the experimental data 
display a smaller band width than DFT by a factor of 1.33, probably due to correlation effects. 
As detailed in the Supplementary Note 2 the sinusoidal bands mainly consist of $d_{yz}$ and $d_{x^2-y^2}$ states, 
whereas the flat band is dominated by states with $d_{zx}$ and $d_{z^2}$ character. 
This observation suggests the presence of a $2\times$ periodicity, which turns the $\overline{X}$ points of the original SBZ 
into $\overline{\Gamma}$ points of the reduced SBZ, as expected for an AFM supercell.
Other features located between $-1$ and $-1.5$\,eV can be interpreted as surface umklapps 
of the Ir bulk band near $\overline{\Gamma}$ that repeat according to the AFM supercell.
We recall here that the ARPES measurements of Fig.\,\ref{Fig:overview}e average over both directions 
parallel and perpendicular to the chains, as a consequence of the domain structure of the system (Fig.\,\ref{Fig:overview}a). 
The inter-chain coupling which results in a $9\times$ magnetic unit cell (see below) 
is expected to be much weaker than the direct intra-chain AFM coupling 
and does not give rise to dispersive features in the ARPES data.

\begin{figure}[t]   
	\includegraphics[width=0.6\columnwidth]{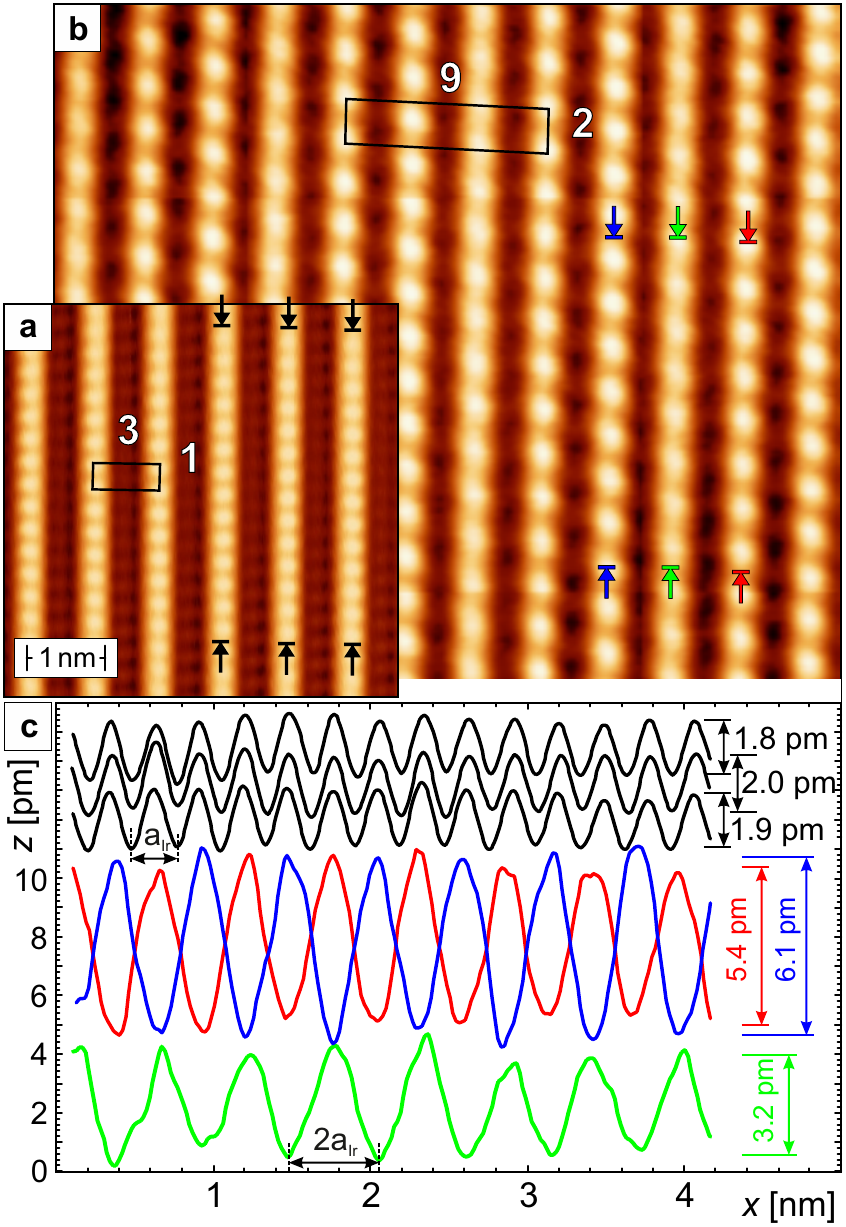}
	\caption{{\bf Atomic resolution scans of MnO$_2$ chains on Ir(001).} 
		{\bf a},~A $(3\,\times\,1)$ structural unit cell is observed with a non-magnetic W tip.  
		{\bf b},~With a Cr-coated W tip the magnetic $(9\,\times\,2)$ unit cell is resolved. 
		{\bf c},~Line profiles drawn along the stripes at the positions indicated by arrows 
		measured with the W (black) and the Cr-coated (colored) probe tip.  
		Spin-resolved line sections differ in periodicity, phase, and amplitude
		from their spin-averaged counterparts (see main text for details). 
		Scan parameters: {\bf a}, $U = -500$\,mV, $I = 3$\,nA;
		{\bf b}, $U = 100$\,mV, $I = 300$\,pA.
				\label{Fig:atomic}}	
\end{figure}
\medskip 
{\bf Spin-polarized scanning tunneling microscopy.} Figure\,\ref{Fig:atomic}a shows an atomic scale STM image 
of MnO$_2$ chains on Ir(001) taken with a non-magnetic W tip. 
The data show a structural $(3\,\times\,1)$ unit cell (black box) 
and nicely reproduce earlier measurements \cite{2016Ferstl}.
The black lines in Fig.\,\ref{Fig:atomic}c show line profiles taken 
along the three adjacent MnO$_2$ chains indicated by arrows in Fig.\,\ref{Fig:atomic}a.  
They exhibit a 
periodicity $(287 \pm 20)$\,pm, 
agreeing well with $a_{\rm Ir}$, i.e., the Mn--Mn inter-atomic distance expected along the chain.
Note, that within the noise level achievable in our setup 
the corrugation amplitude of $(1.9 \pm 0.1)$\,pm remains constant.  

As we will describe in the following, our spin-polarized STM experiments exhibit some additional contrasts 
which allow to elucidate the spin structure of the MnO$_2$ chains on Ir(001).  
Fig.\,\ref{Fig:atomic}b shows an SP-STM image scanned with an in-plane sensitive Cr-coated W tip.  
Comparison with the spin-averaged data presented in Fig.\,\ref{Fig:atomic}a reveals two qualitative differences:
(i) The periodicity measured with magnetic tips along the chains is longer and 
(ii) the contrast observed on different MnO$_2$ chains is not constant 
but becomes significantly smaller for every third chain.  
Again we analyzed line profiles taken along three adjacent MnO$_2$ chains 
in between the colored arrows in Fig.\,\ref{Fig:atomic}b. 
These data, which are plotted in the bottom part of Fig.\,\ref{Fig:atomic}c, 
immediately illustrates a doubling of the periodicity, $2a_{\rm Ir}$.
This SP-STM contrast is characteristic for alternating spins \cite{KFB2005}
and consistent with the proposed AFM Mn--Mn coupling along the chains \cite{2016Ferstl}.   
Furthermore, the spin-polarized data reveal a systematic variation of the corrugation. 
For example, the blue line trace exhibits a corrugation of 6.1\,pm, 
in contrast to 3.2\,pm (green) and 5.4\,pm (red) for the two adjacent MnO$_2$ chains.  
Finally, there is also a distinct phase relation between the chains. 
Comparing the three colored traces plotted in Fig.\,\ref{Fig:atomic}c a $\uppi$ phase shift 
between the blue and the green trace becomes apparent, 
whereas no phase shift occurs between the green and the red trace.  

\begin{figure}[t]   
	\includegraphics[width=0.7\columnwidth]{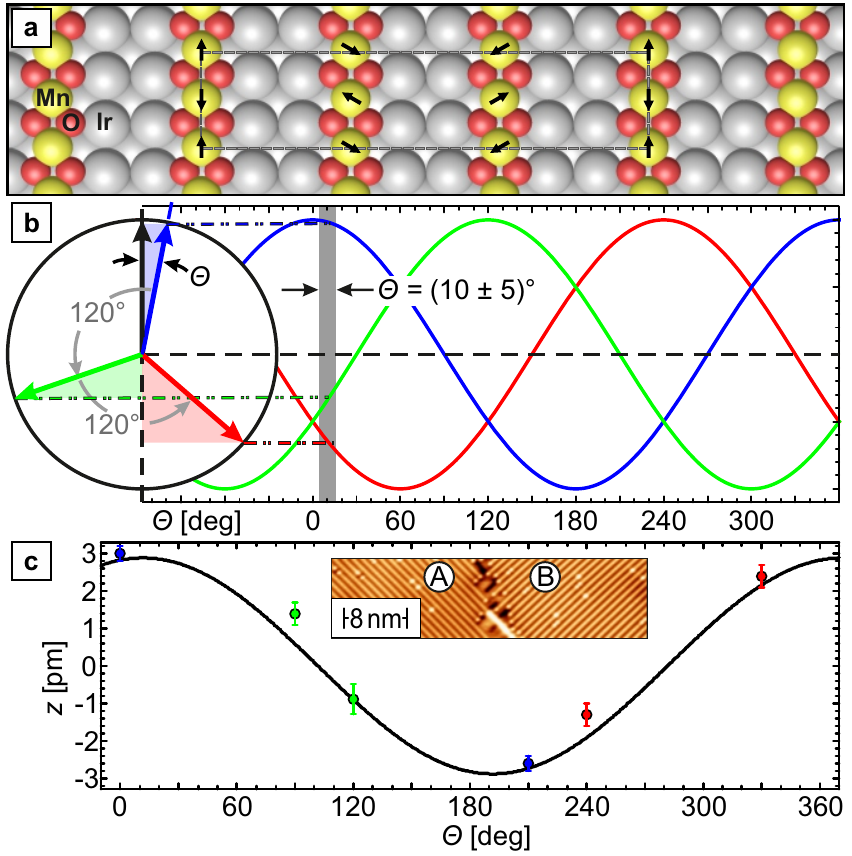}
	\caption{{\bf Interpretation of SP-STM results by a chiral inter-chain exchange.  } 
		{\bf a},~Sche\-matic model of the chiral $(9\,\times\,2)$ spin structure of MnO$_2$/Ir(001).
		{\bf b},~Sketch of the SP-STM signal expected on a sample with three domains 
		rotated by $120^{\circ}$ to another (see text and Eq.\,\ref{SPcurrent} for details).  
		The corrugations data of Fig.\,\ref{Fig:atomic}b are best fit with an angle $\theta = (10 \pm 5)^{\circ}$ 
		between tip magnetization (black) and the nearest domain (blue). 	
		{\bf c},~Corrugation values determined by fitting line profiles measured 
		on two adjacent domains A and B (inset) as measured with an in-plane tip. 
		The error bar represents the residual sum of squares. 
		The black curve represents the cosine expected for an in-plane rotating spin spiral.
		\label{Fig:Scheme}}
\end{figure}
\medskip 
{\bf Modeling the magnetic contrast.} The SP-STM contrast observed on MnO$_2$ chains
can semi-quantitatively be understood by assuming an AFM coupling along the chain 
and a chiral $120^{\circ}$ coupling between adjacent chains. 
This spin configuration which leads to a $(9\,\times\,2)$ magnetic unit cell 
is schematically sketched in Fig.\ \ref{Fig:Scheme}a.
As mentioned above the AFM Mn--Mn intra-chain coupling 
can directly be concluded from the doubling of the periodicity in SP-STM  
as compared to spin-averaged data ($a_{\rm Ir}$) (cf.\ Fig.\,\ref{Fig:atomic}c). 
To also explain the corrugation amplitudes and their phase 
we need to consider that the magnetic corrugation in SP-STM, $\Delta z_{\rm SP}$, 
depends on the cosine of the angle $\theta$ included between 
the magnetization directions of the tip 
and the sample, 
\begin{equation}
\Delta z_{\rm SP} \propto  P_{\rm t} \cdot P_{\rm s}
			\cdot \cos \theta, 
\label{SPcurrent}
\end{equation}
with $P_{\rm t}$ and $P_{\rm s}$ being the spin polarization of tip and sample, respectively. 
The expected magnetic contrast can be deduced from the scheme in Fig.\,\ref{Fig:Scheme}b. 
It represents three sample magnetization directions which are rotated 
by $120^{\circ}$ to another as symbolized by colored arrows.  
According to Eq.\,\ref{SPcurrent}, $\Delta z_{\rm SP}$ is given by the projection 
of the sample magnetization 
onto the tip magnetization. 
Therefore, the maximum $\Delta z_{\rm SP}$ is expected for a sample magnetization 
which is almost collinear to the tip magnetization (represented by the black arrow). 
As symbolized by the lightly colored triangle this condition 
is fulfilled for the blue arrow in Fig.\,\ref{Fig:Scheme}b (offset by angle $\theta$). 
In this situation it is unavoidable that the projection of the other two arrows 
points into the direction opposite to the black arrow.  
This can also be verified by inspecting the right part of Fig.\,\ref{Fig:Scheme}b 
where we plot three cosine functions shifted by $120^{\circ}$.  

In other words, from the fundamental principles of SP-STM it follows 
that (i) whenever we obtain a large contrast on one AFM spin chain 
the other two chains with spin quantization axes rotated by $\pm 120^{\circ}$ 
must exhibit a magnetic corrugation which is phase-shifted with respect to the high-contrast row.  
Furthermore, (ii) even if $\theta$ is relatively small,
one of the $\pm 120^{\circ}$-rows exhibits a much lower magnetic contrast 
since 
the $\cos \theta$ term in Eq.\,\ref{SPcurrent} is close to zero.  
As marked by a grey box in Fig.\,\ref{Fig:Scheme}b, 
the corrugations measured 
in Fig.\,\ref{Fig:atomic}b 
can nicely be explained by a tip which is rotated by $\theta = (10 \pm 5)^{\circ}$
with respect to the (blue) domain. 

As discussed in detail in the Supplementary Note 3 we have performed 
various SP-STM measurement to identify the spin orientation of the MnO$_2$ chains.
Fig.\,\ref{Fig:Scheme}c shows the magnetic corrugation measured 
on the three MnO$_2$ chains (blue, red, and green) of two domains (A and B; see inset), respectively, 
which 
are rotated by $90^{\circ}$ with respect to another.  
The black line is the corrugation expected for a spin spiral rotating in the surface plane.  
The very good agreement with our experimental data suggests an inter-chain coupling 
characterized by an in-plane $120^{\circ}$ rotation of the azimuthal spin orientation.  
In order to verify if this spin order is indeed chiral we determined the rotational sense 
of seven separate MnO$_2$ domains (see Supplementary Note 4). 
Indeed, our SP-STM measurements show that all domains exhibit the same rotational sense, 
a result which is highly unlikely under non-chiral conditions ($< 2 \%$). 

\medskip 
{\bf Density-functional theory calculations.} 
To obtain some insights into the origin of the observed magnetic structures we performed DFT calculations
(see Supplementary Note 5 for details). 
The preferred magnetic ordering along the chains was found to be AFM, 
in agreement with earlier calculations \cite{2016Ferstl} and our experimental results (see Fig.\,\ref{Fig:atomic}b). 
Whereas a weak AFM inter-chain coupling of $0.4$\,meV per Mn atom was found previously \cite{2016Ferstl}, 
our calculations performed at a much denser  $\mathbf{k}$-point sampling 
(using a $24 \times 36$ Monkhorst-Pack grid) leads to a weak FM coupling of $1.7$\,meV. 
We calculated flat spin spirals with various wave vectors $q$, 
where the FM (AFM) state corresponds to $q = 0$ ($q = 0.5$) in units of $2 \uppi / 3 a_{\rm Ir}$.
\begin{figure}[t]   
        \includegraphics[width=0.65\columnwidth]{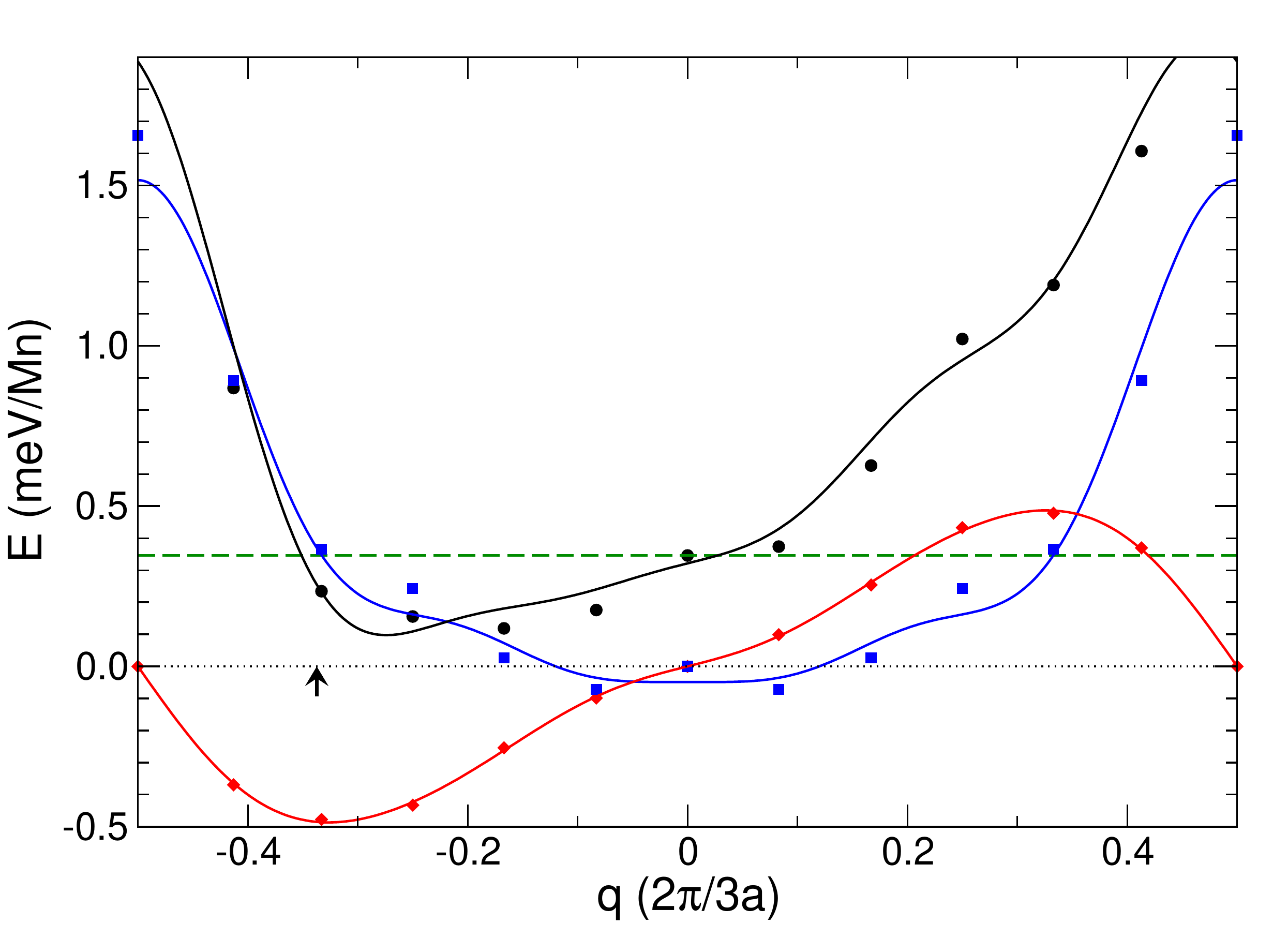}
        \caption{{\bf Results of DFT calculations.} 
		Total energy of AFM ordered MnO$_2$ chains on Ir(001) 
		where the spin directions from chain to chain rotate by an angle of $3a_{\rm Ir}$. 
		The spins form a cycloidal spin spiral propagating perpendicular to the chains 
		with the rotation axis parallel to the chain direction. 
		The contribution from symmetric exchange interaction is shown by blue squares, 
		from the DMI by red diamonds, and from the magneto-crystalline anisotropy by the dashed line. 
		The sum of these terms is marked by black circles. 
		The blue line is a fit of the symmetric exchange interaction with $\cos(2 n \uppi q)$, 
		likewise the DMI was fitted (red line) with $\sin(2 n \uppi q)$, both up to $n = 4$. 
	\label{Fig:DFT}}
\end{figure}
From our spin spiral calculations, Fig.~\ref{Fig:DFT}, we can see that symmetric (Heisenberg-type; blue) 
exchange interactions lead to a flat dispersion, without any minimum at finite $q$. 
Our results show that the DMI is largest for a spin spiral with $q = 1/3$, 
i.e., the modulation vector found experimentally, lowering the total energy by about $0.3$\,meV. 
We have to note, however, that the theoretically obtained Dzyaloshinskii vector, $\mathbf D$, 
points along the chain direction, whereas experiments suggest an in-plane spin spiral, 
corresponding to $\mathbf D$ along the surface normal. 
To explain the experimentally observed unique rotation sense, 
there must be a significant out-of-plane component of $\mathbf D$, 
e.g.\ due to a structural distortion that removes the $[\bar 1 1 0]$ mirror plane. 
A similar mechanism has recently been shown to exist for zigzag Co/Ir(001) \cite{2015Dupe}.  
Indeed, some hints of a potential distortion of the $(3\,\times\,1)$ structural unit cell 
can not only be recognized in our data (Fig.\,\ref{Fig:atomic}a), 
but also in the data published by Ferstl and co-workers (see Fig.\,1c and Fig.\,S2 in Ref.\,\onlinecite{2016Ferstl} 
and the Supplementary Note 6 of this article). 
In either case atomic resolution data recorded on magnetic TMO chains with non-magnetic tips 
show some oblique distortion of the expected rectangular surface unit cell. 
Although this does not directly lead to a non-vanishing effective perpendicular $\mathbf D$,
it indicates that some structural details or relaxation effects due to the finite size of structural domain 
still need to be resolved (see Supplementary Note 5).

Recent research indicates that spin-orbit coupling supports an effective spin transfer 
torque\cite{Jonietz2010,Emori2013,Ryu2014} which is particularly important in applications.   
Different mechanisms have been proposed to explain the relatively low current thresholds 
necessary to drive skyrmions or chiral domain walls, including inhomogeneous spin currents \cite{Jonietz2010}, 
Rashba fields, the spin Hall effect \cite{Emori2013}, the DMI, or a combination of the latter two \cite{Ryu2014}. 
We speculate that a DME-RKKY interaction-induced indirect chiral magnetic exchange 
may also lead to an extreme reduction of the required current density in layered magnetic structures.  
It remains to be investigated whether a chiral magnetic interlayer coupling 
as it has been observed in Dy/Y superlattices  \cite{GCY2008,GLC2010} can also be found 
in other material combinations with strongly spin-orbit coupled non-magnetic spacer layers.  
In more general terms, DME-RKKY interaction may give rise to rather exotic phenomena, 
such as chiral spin-liquid states in spin ice systems \cite{Machida2009,Flint2013} 
or the emergence of new quasipar\-ticles due to the trapping of single electrons 
in self-induced skyrmion spin textures \cite{Brey2017}.  

In summary, we have investigated the intra- and inter-chain magnetic coupling 
of the quasi one-dimensional system of structurally $(3\,\times\,1)$-ordered MnO$_2$ on Ir(001) 
by spin-polarized scanning tunneling microscopy, angle-resolved photoemission, and density functional theory. 
Both experimental methods confirm an antiferromagnetic order along the chains, as predicted earlier \cite{2016Ferstl}. 
In addition, spin-polarized scanning tunneling microscopy reveals a chiral $120^{\circ}$ rotation 
between adjacent MnO$_2$ chains, resulting in a $(9\,\times\,2)$ magnetic unit cell.  
Density functional theory finds that a Dzyaloshinskii-Moriya type enhancement of the RKKY interaction 
indeed leads to chiral interchain coupling with a periodicity in agreement with experiment.
However, the orientation of the Dzyaloshinskii vector $\mathbf D$ remains to be clarified.  
Whereas experimental results suggest a perpendicular $\mathbf D$, 
theory predicts a $\mathbf D$ vector oriented along the chains.

\section*{Methods}
\vspace{-0.5cm}
{\bf Sample preparation.}
Sample preparation procedures closely follow published recipes \cite{2016Ferstl}.
Initial Ir(001) preparation comprises cycles of ion-sputtering (1\,keV, Ar$^+$, $\approx 2\,\mu$A) 
followed by annealing to 1400\,K in an oxygen atmosphere.
The pressure gauge indicates a background pressure $p_{\rm O_2} \approx 1 \times 10^{-8}$\,mbar,
but since the gas nozzle is located a few cm above the sample 
the local oxygen pressure is assumed to be about two orders of magnitude higher.
We obtain the $(5\,\times\,1)$ reconstruction characteristic for clean Ir(100) 
by an annealing cycle without oxygen \cite{SMH2002}.
Oxidizing this surface again in $p_{\rm O_2} \approx 1 \times 10^{-8}$\,mbar at $T_{\rm S} \approx 850$\,K 
leads to the oxygen-terminated Ir(100)-$(2\,\times\,1)$ reconstruction \citep{1976Rhodin,2000Johnson,2016Ferstl}.
It served as a substrate for the deposition of 0.33 monolayers (ML) of Mn at room temperature, 
followed by final annealing ($T_{\rm S} \approx 1050$\,K) under oxygen atmosphere.
 
{\bf Scanning tunneling microscopy.}
STM experiments were performed in a two-chamber ultra-high vacuum (UHV) system 
(base pressure $p \leq 5 \times 10^{-11}$ mbar) equipped with a home-built 
low-temperature scanning tunneling microscope (LT-STM) (operation temperature $T = 5.5$\,K).
We used electro-chemically etched polycrystalline W tips which were flashed by electron bombardment 
and coated with Fe or Cr for SP-STM measurements \cite{Bode2003}.
 
{\bf Angular-resolved photoemission spectroscopy (ARPES). }
ARPES data were acquired at 130 and 150\,eV photon energy at the VUV photoemission beamline (Elettra, Trieste). 
These photon energies are close to the Cooper minimum of the Ir $5d$ photoemission (PE) cross section\,\cite{YEH19851} 
and enhances the PE signal of Mn $3d$ states (chains) with respect to the overlapping Ir $5d$ states (substrate). 
The spot of the synchrotron light on the sample (500 $\mu$m $\times$ 200 $\mu$m) 
is much larger than the typical size of domains with parallel MnO$_2$ chains. 
Thus, ARPES provides a space-averaged signal over the two perpendicular orientations of the MnO$_2$ chains. 
The energy and momentum resolutions were set to 15\,meV and 0.02 \AA$^{-1}$, respectively.

{\bf DFT calculations}
Non-collinear DFT calculations were performed using the full-potential linearized augmented plane wave method 
as implemented in the {\sc Fleur} code~\cite{Kurz:04.1}.
We set up a seven layer film in a $(3 \times 2)$ unit cell as described in Ref.\,\onlinecite{2016Ferstl},
using the local density approximation~\cite{Vosko:80.1} 
with Hubbard $U$ corrections~\cite{Shick:99.1} on the Mn $d$ states ($U = 2.7$\,eV, $J = 1.2$\,eV).
We confirmed that these values put the Mn $d$ states about $2.2$\,eV below the Fermi level,
in good agreement with the ARPES data presented in Fig.\,\ref{Fig:overview}d.
We used the generalized Bloch theorem to calculate the spin spiral structures and included spin-orbit coupling 
in first order perturbation theory to estimate the strength of the DMI in this system~\cite{Heide:09.1}. \vspace{-0.5cm}

\section*{Data availability} 
\vspace{-0.5cm}The data that support these findings of this study are available on request 
from M.S.\ (STM), P.M.\ (ARPES), and G.B.\ (theory).  \vspace{-0.5cm}

\section*{Acknowledgments} 
\vspace{-0.5cm}Experimental work was supported by DFG through FOR 1700 (project E6), SPP 2137 ``Skyrmionics'' (BO 1468/26-1), 
and by the Dresden-W{\"u}rzburg Center for Topological Quantum Matter Research (ct.qmat).
A.K.K.\ acknowledges receipt of a fellowship from the ICTP-TRIL Programme, Trieste, Italy.

\section*{Author contributions}
\vspace{-0.5cm}M.S., R.C., M.V., and J.K.\ performed and analyzed STM measurements. 
P.M., P.M.S., A.K.K., and C.C.\ performed ARPES experiments and analyzed the data.  
G.B., A.B., and S.B.\ provided the theoretical framework. 
All authors discussed the results and contributed in writing the manuscript.

\section*{Competing Interests} 
\vspace{-0.5cm}The authors declare no competing interests.

\bibliography{MnO_03}

\end{document}